\documentclass[aps,prd,twocolumn,nofootinbib,showpacs,superscriptaddress]{revtex4-1}
\usepackage[frenchb,english]{babel}
\usepackage[latin1]{inputenc}
\usepackage{enumerate}
\usepackage{amsmath}
\usepackage{amsthm}
\usepackage{amssymb}
\usepackage{graphicx}
\usepackage{floatflt}
\usepackage{fancybox}
\usepackage{stmaryrd}
\usepackage{appendix} 
\usepackage{enumitem} 
\usepackage{subfigure}
\usepackage[b]{esvect}
\usepackage{hyperref} 
\usepackage{color}
\usepackage{sidecap}

\newcommand{\om}{\omega}

\newcommand{\n}{{w}} 
\newcommand{\al}{\alpha} 

\renewcommand{\vec}[1]{\vv{#1}}
\newcommand{\p}{\partial}

\newcommand{\be} {\begin{equation}}
\newcommand{\ee} {\end{equation}}
\newcommand{\bsub}[1]{\begin{subequations} \label{#1} \begin{eqnarray}}
\newcommand{\esub}{\end{eqnarray} \end{subequations}}
\newcommand{\bea}{\begin{eqnarray}}
\newcommand{\eea}{\end{eqnarray}}
\newcommand{\bi} {\begin{itemize}}
\newcommand{\ei} {\end{itemize}}
\newcommand{\ben} {\begin{enumerate}}
\newcommand{\een} {\end{enumerate}}
\newcommand{\bmat} {\begin{pmatrix}}
\newcommand{\emat} {\end{pmatrix}} 
\newcommand{\bal} {\begin{aligned}}
\newcommand{\eal} {\end{aligned}}
\newcommand{\btab}{\begin{tabular}}
\newcommand{\etab}{\end{tabular}}

\newcommand{\eq}[1]{Eq.~\eqref{#1}}
\begin{document}
\selectlanguage{english}
\title{An acoustic probe for quantum vorticity in Bose--Einstein condensates}
\author{Antonin Coutant}
\email{antonin.coutant@aei.mpg.de}
\affiliation{Max Planck Institute for Gravitational Physics, Albert Einstein Institute, Am M\"uhlenberg 1, 14476 Golm, Germany.}
\author{Silke Weinfurtner}
\email{silke.weinfurtner@nottingham.co.uk}
\affiliation{School of Mathematical Sciences, University of Nottingham, University Park, Nottingham, NG7 2RD, UK} \affiliation{Perimeter Institute for Theoretical Physics, Waterloo, Ontario, Canada N2L 2Y5}
\begin{abstract}
We investigate the deformation of wavefronts of sound waves in rotating Bose--Einstein condensates. In irrational fluid flows Berry et al. identified this kind of deformation as the hydrodynamic analogue of the Aharonov--Bohm effect. We study this effect in Bose--Einstein condensates and obtain the Aharonov--Bohm phase shift at all wavelengths. We show that this deformation of wave fronts is seen in both phase \emph{and} density fluctuations. For wavelengths larger than the healing length, the phase fluctuations experience a phase shift of the order of $2\pi$ times the winding number. We also consider lattices of vortices. If the angular momentum of the vortices are aligned, the total phase shift  is $2\pi$ times the number of vortices in the condensate. Because of this behaviour the hydrodynamic Aharonov--Bohm can be utilized as a probe for quantum vorticity, whose experimental realization could offers an alternative route to investigate quantum turbulence in the laboratory.
\end{abstract}
\pacs{62.60.+v,67.25.dk,67.85.Jk,67.85.Hj} 
\maketitle
In 1980 Berry, Chambers, Large and Upstill~\cite{Berry80} realized that surface waves scattered on a vortex pick up a geometric phase shift. This geometric phase shift turns out to be the exact analogue of the Aharonov--Bohm phase shift~\cite{Aharonov59}. Moreover, the absence of gauge symmetry in the hydrodynamic analogue allows for new features that are unobservable in the quantum mechanical problem. In a fluid, the wave undergoes a wavefront dislocation~\cite{Berry81} that is, entirely new wavefronts come out of the vortex core. This hydrodynamic Aharonov--Bohm effect has been considered by several authors, for water waves~\cite{Coste99,Coste99b,Vivanco99} or for superfluid helium~\cite{Lund95,Davidowitz97}. The main difference between classical and quantum fluids, is that the vortex circulation is quantized. According to Berry's classification~\cite{Berry95}, there are four types of Aharonov--Bohm effect, depending whether the wave and/or the flux (here the vortex circulation) is quantum or classical. Here we investigate the Aharonov--Bohm effect on sound waves (classical) scattered on Bose--Einstein condensates vortices (quantum). 

Wavefront dislocations have been used theoretically and experimentally in Bose--Einstein condensates to detect vortices~\cite{Chevy01} and their solitonic version~\cite{Donadello14}. This dislocation arises from a geometric phase in the condensate wave function. Here instead, we shall investigate how vortices affect plane wave perturbations. The aim of the present paper is twofold. We present a detailed analysis of the analogue Aharonov--Bohm effect, which includes dispersive effects, and compare it with the dislocation of the condensate wave function itself. The hydrodynamic Aharonov--Bohm effect in a Bose--Einstein condensate has so far not been considered. As we show, for wavelengths of the order of the healing length, the induced phase shift is of order unity and should therefore be experimentally measurable. This effect also cumulates when considering a lattice of vortices, which is the stable configuration of rapidly rotating condensates~\cite{Fetter09}. This allows us to probe vortices using sound waves, and obtain not only the total angular momentum, but also its orientation, and hence offers an alternative method to explore quantum turbulence in the laboratory. \\

\emph{{Quantum vorticity.---}} 
In the Bogoliubov approximation, a Bose--Einstein condensate is described by its condensate wave function $\Psi$~\cite{Dalfovo99}. Its modulus gives the atom density, and its phase the velocity field, that is 
\be \label{fluct_modes}
\Psi(t,\vec r) = \sqrt{n(\vec r)} \exp\left({-i \mu t + i \Theta(\vec r)}\right). 
\ee
Here $\mu$ is the chemical potential, and $n(\vec r)$ the local number density. In units where $\hbar = 1$, the velocity field is given by $\vec v = \vec \nabla \Theta/M$, where $M$ is the mass of the atoms of the condensate. In a condensate, the flow is always irrotational, as it is by definition a gradient. However, wherever the density vanishes, the phase becomes ill-defined, and the velocity field can acquire a nonzero circulation. This give rise to vortices in the condensate (see~\cite{Fetter09} for a review). If a single vortex is present, the condensate wave function reads 
\be \label{vortex_wf}
\Psi(t,\vec r) = \sqrt{n(r)} e^{-i \mu t + i \n \theta},
\ee
where we used cylindrical coordinates $(r,\theta,z)$. Because the condensate wave function must be single valued, $\n$ is an integer. This means the circulation of the velocity field is quantized and given by  
\be \label{JBEC}
C_\n = 2\xi c \n, 
\ee
where $c$ is the speed of sound in the condensate and $\xi = (2Mc)^{-1}$ the healing length. Around a vortex, the density is essentially constant $n(r) \sim n_0$, except within a region of size $\sim \xi$, which corresponds to the core of the vortex~\cite{Fetter65}. Inside that region, the density decreases until it vanishes at $r=0$, the center of the vortex. Moreover, in a real condensate, quantum depletion further dresses the flow near the core of the vortex. The non-condensed part of the atoms induces a flow that is no longer restricted to be curl-free, and one can show that close to the center, the flow essentially becomes that of a solid rotation~\cite{Fetter72}. In the sequel, we shall consider perturbation far from the vortex core, so that we can safely approximate the density by $n(r) \sim n_0$ and the velocity flow by its superfluid part 
\be \label{vortexv}
\vec v = \frac{C_\n}r \vec{e_\theta} , 
\ee
which is the expression for an irrotational vortex ($\vec{e_\theta}$ is the unit angular vector). When two such condensates interfere, the pattern shows a wavefront dislocation on each vortex core~\cite{Chevy01,Donadello14}. The number of new wave fronts is precisely given by $\n$. As we shall see, perturbations around the condensate undergo a similar effect. However, depending on the wavelength, the geometric phase will smoothly interpolate between 0, and $2\pi \n$, i.e. $\n$ dislocated fronts. 
\medskip

\emph{Sound waves.---}
It is well-known that excited states of a condensate can be obtained from the eigen-modes of the linearized Gross-Pitaevskii equation~\cite{Dalfovo99}. For this we assume a condensate wave function of the form 
\be
\Psi(t,\vec r) = \psi_0(t,\vec r) \left(1 + u_\om(\vec r) e^{-i \om t} + v_\om^*(\vec r) e^{i \om t}\right),
\ee
where $\psi_0$ is given by \eq{vortex_wf}. The fluctuations modes $u_\om$, $v_\om$ then obey the Bogoliubov-deGennes equation~\footnote{The operator $\{\vec v \cdot \vec \nabla\}$ should be understood as $(\vec v \cdot \vec \nabla + \vec \nabla \cdot \vec v)/2$.} 
\bsub{BdG_eq} 
\om u_\om &=& \left(-\frac{\Delta}{2M} + i \{\vec v \cdot \vec \nabla\} + Mc^2\right) u_\om + Mc^2 v_\om , \qquad \\
-\om v_\om &=& \left(-\frac{\Delta}{2M} - i \{\vec v \cdot \vec \nabla\} + Mc^2\right) v_\om + Mc^2 u_\om . \qquad
\esub
In the absence of a flow, the modes can be decomposed in plane waves $u_\om(\vec r) = U_k e^{i \vec k \cdot \vec r}$ of momentum $\vec k$. The above equations then impose the Bogoliubov dispersion relation 
\be \label{Bogo_disp}
\om_k^2 = c^2 (k^2 + \xi^2 k^4) = F(k^2). 
\ee
We point out that we do not decompose $u_\om$ and $v_\om$ in circular waves of fixed azimutal number (eigen-modes of $i\p_\theta$) as is usually done~\cite{Pitaevskii61,Fetter65b}. This is because we want to see how \emph{plane wave} excitations are affected by the vortex. We shall now proceed in two steps. We first solve the \eq{BdG_eq} to obtain the long distance pattern of a plane wave on the vortex, and then discuss its physical consequences. 
\medskip

\emph{Wavefronts in the acoustic ray approximation.---} 
We now solve \eq{BdG_eq} using a ray, or WKB approximation. We assume the wave is propagating perpendicularly to the vortex axis. For this, we consider a solution of \eq{BdG_eq} independent of $z$ and of the form 
\be \label{WKBmode}
u_\om = U_k(\vec r) e^{i S(r,\theta)} \quad \textrm{and} \quad v_\om = V_k(\vec r) e^{i S(r,\theta)},
\ee
where $U_k$ and $V_k$ are (real) slowly varying amplitudes and $S(r,\theta)$ the wave action. Putting this \emph{ansatz} in \eq{BdG_eq}, and performing a gradient expansion~\footnote{Such a WKB method applied higher derivatives equations is explain in great details in~\cite{Coutant11}, see in particular Appendix A.}, we find that $S$ is solution of the Hamilton-Jacobi equation 
\be \label{HJ}
(\om - \vec v \cdot \vec \nabla S)^2 = F\left((\vec \nabla S)^2\right). 
\ee
In general this equation is solved by separation of variables, i.e. assuming a single azimutal number. Since we want to consider an incoming plane wave, we shall instead obtain an approximate solution by treating the velocity field as a perturbation. Since $\vec v \to 0$ at infinity, this will become valid far enough from the center of the vortex. As mentioned above, in the absence of velocity field, we simply have $\vec \nabla S = \vec k$. In the presence of the vortex, we assume that the perturbation on the wave action is proportional to $\vec v$ 
\be \label{WKBaction}
\vec \nabla S = \vec{k} + \gamma \vec v,  
\ee 
where $\gamma$ is position independent. Since $\vec v \to 0$ at infinity, this equation means that there is an incoming plane wave of momentum $\vec k$ on the vortex. Assuming $\gamma \vec v$ is a small correction, we solve the Hamilton-Jacobi \eq{HJ} perturbatively. Expanding to first order, the derivative of $F$ gives the Bogoliubov group velocity via $v_g = k \p_k F/\om$. Hence, we obtain the expression of $\gamma = - \frac{k}{v_g}$ and deduce the wave action $S(r,\theta)$ that governs the phase of WKB modes in \eq{WKBmode}
\be \label{Shydro}
S(r,\theta) = \vec{k} \cdot \vec r - \al \theta , 
\ee
where 
\be \label{ABparam}
\al = \frac{kC_\n}{v_g} .
\ee
A key point is that the discontinuity in $\theta$ must be the half line directed by $\vec k$ (we shall briefly return to that below). When the wave comes from the left, $\vec k \cdot \vec r = r \cos(\theta)$ and $\theta \in ]0;2\pi[$. The key point of this result is that despite the fact that $\vec v \to 0$ far from the vortex, there is a nonzero imprint on the profile of the modes, that is, they do not reduce to a plane wave. This is due to the fact that $\vec v$ falls off as $O(1/r)$ at large $r$, something reminiscent to long range potentials in scattering theory~\cite{Newton}. The parameter $\al$ governs a long range effect induced by the velocity profile~\eqref{vortexv}. In addition, under the same approximations, \eq{BdG_eq} gives a linear relation between the amplitudes $V_k = D_k U_k$, with 
\be
D_k = \left(2 \xi k \sqrt{1+ \xi^2 k^2} - 2\xi^2 k^2 - 1 \right). 
\ee

\emph{Analogue Aharonov--Bohm effect in a BEC.---} 
When using \eq{Shydro} for WKB modes \eqref{WKBmode}, we see that a plane wave scattered on a vortex is distorted by the second term of \eq{Shydro}. This is the hydrodynamic analogue of the Aharonov--Bohm effect. When travelling around the vortex, for $\theta$ from 0 to $2\pi$, we accumulate a geometric phase of $2\pi \al$. In the original Aharonov--Bohm effect, the wave function is \emph{not} a direct observable, and the electromagnetic gauge symmetry only allows us to observe $\al$ modulo an integer~\cite{Berry80}. Here, the perturbation field is a direct observable and extra features of the wave are described by \eq{Shydro}. First, there are wave front dislocations governed by the integer part of $\al$, i.e. there is $[\al]$ extra wave fronts coming out from the vortex core. This is very similar to the dislocation that one observes on the condensate wave function itself. Second, there is a line of discontinuity on $\theta = 0$ when $\al$ is not an integer. This feature cannot be seen on the condensate wave function, but only in the perturbations. 

\begin{figure}[!t]
\begin{center}
\includegraphics[width=\columnwidth]{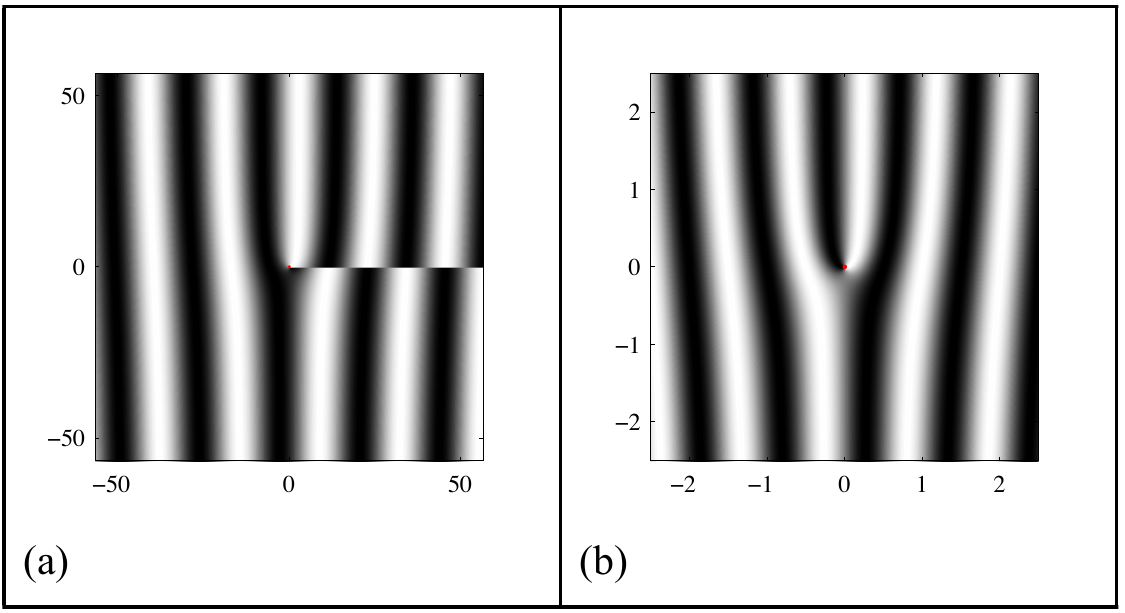}
\caption{Hydrodynamic Aharonov--Bohm effect on a singly quantized vortex with $w=+1$: amplitude of $\delta \theta$ \emph{or} $\delta n$ of \eq{Observ}. Scale in the $x$ and $y$ directions is in units of the healing length. Red dots indicate the position of a vortex turning anti-clockwise. (a) $\xi k = 1/(4\pi)$ thus $\al\sim1.6$ and (b) $\xi k = 1$  thus $\al\sim1$. Depending on $\al$ of \eq{ABparam}, one can observe a dislocation and\,/\,or a discontinuity in the forward direction of the plane wave.
\label{fig-single-vortex}} \vspace{-0.5cm}
\end{center}
\end{figure}

Using the Bogoliubov dispersion relation, we compute the group velocity and use \eq{JBEC} to deduce the expression of the Aharonov--Bohm parameter \eqref{ABparam}. 
\be
\al = 2 \xi k w \frac{\sqrt{1+\xi^2 k^2}}{1+2\xi^2 k^2}.
\ee
It is instructive to explore this result in two opposite regimes: the long and short wavelength limit. 
\be
\al \sim \left\{\bal  2 \n \xi k &\quad \textrm{if} \quad \xi k \ll 1, \\
 \n &\quad \textrm{if} \quad \xi k \gg 1.
 \eal\right. 
\ee
In the long wavelength limit, the phase shift vanishes. This is because long wavelength excitations cannot be distinguished from the condensed part itself, i.e. they rotate the same way, and therefore, the relative phase disappears. For short wavelengths, the phase shift reduces to $2\pi$ times the winding number of the vortex. This is due to the fact that wavelengths shorter than the healing length correspond to single particles, barely seeing the condensate. Hence, they propagate without any extra phase, and the geometric phase of relative fluctuations is simply that of the condensate wave function. Additionally, this geometric phase affects both density and phase fluctuations of the condensate. Indeed, from \eq{fluct_modes}, density fluctuations $\delta n$ are given by the real part of perturbations and phase fluctuations $\delta \theta$ are given by the imaginary part.  By choosing the phase convention such that $U_k$ is real, we obtain 
\bsub{Observ}
\delta \theta &=& (1-D_k) U_k \sin\left(- \om t + \vec{k} \cdot \vec r - \al \theta \right), \\
\frac{\delta n}{2n_0} &=& (1+D_k) U_k \cos\left(- \om t + \vec{k} \cdot \vec r - \al \theta \right).
\esub
Therefore, both observables show the same geometric phase shift, that is the analogue Aharonov--Bohm effect. The amplitudes of the perturbations are both governed by $U_k$. This relation between $\delta n$ and $\delta \theta$ is a direct consequence of the continuity equation. 
\medskip

\emph{Beyond WKB: the scattered wave.---} 
Since $\vec v$ and its derivative vanish for $r\to \infty$, the WKB approximation becomes exact at long distances. In fact, WKB modes gives the incoming part of scattering states. A complete scattering state includes also a scattered wave and has the asymptotic form 
\be \label{ScatMode}
u_\om(r,\theta) \sim e^{ i \vec k \cdot \vec r - i \al \theta} + f_k(\theta) \frac{e^{i k r}}{\sqrt{r}}. 
\ee
This is interpreted in a standard way~\cite{Newton}. The first term represents the incoming plane wave while the second one give the scattered wave whose amplitude $f_k(\theta)$ varies for different directions. \eq{ScatMode} shows that the incoming part of stationary states does not reduce to a plane wave, but picks up the geometric phase governed by \eq{Shydro}, something that was unnoticed in previous studies of scattering on hydrodynamic vortices~\cite{Fetter64,Fetter65b,Dolan11,Dolan12}. The second term of \eqref{ScatMode} interfere with the incoming part to modify the wave profile. By comparing our Fig.~\ref{fig-single-vortex} with that of e.g.~\cite{Coste99}, we see that this mildly affects the main features described above. 

Moreover, when $\al$ is not an integer, the expression \eqref{Shydro} is only valid away from the line $\theta = 0$, where there is a discontinuity of the wave fronts (see Fig.~\ref{fig-single-vortex}~(a)). When including the scattered part, the scattering amplitude diverges in the forward direction, i.e. $f_k(\theta \to 0) \to \infty$. This shows that the WKB approximation breaks down in the forward direction. When considering exact solutions, the asymptotic \eqref{ScatMode} is valid only outside a small angular region of width $\Delta \theta = O((kr)^{-1/2})$ where the mode smoothly interpolates between the WKB wave fronts on each side~\cite{Berry80}. 
\medskip

\begin{figure}[!h]
\begin{center}
\includegraphics[width=\columnwidth]{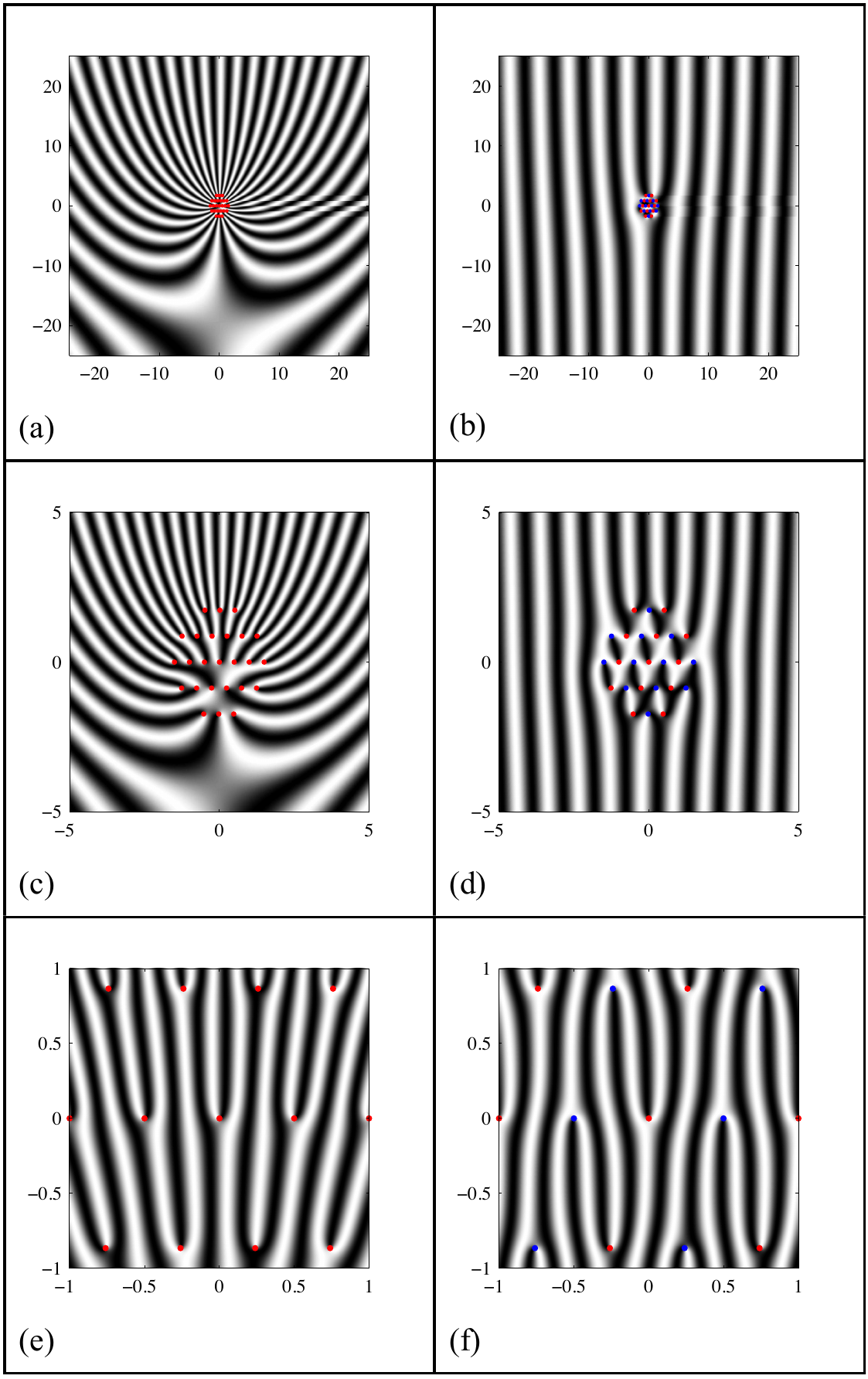}
\caption{Hydrodynamic Aharonov--Bohm effect on a hexagonal vortex lattice: amplitude of $\delta \theta$ \emph{or} $\delta n$ on 25 vortices arranged to a hexagonal lattice. Scales in the $x$ and $y$ direction is in units of the healing length, and vortices are separated by approximately $2\xi$. Red vortices are turning anti-clockwise, and blue vortices clockwise. (a,b) $\xi k = 1/5$, (c,d) $\xi k = 1$, (e,f) $\xi k = 5$. (Notice on the right panels the total circulation is one anti-clockwise vortex.) \vspace{-0.5cm}}
\label{front-fig}
\end{center}
\end{figure}

\emph{Multiple vortices.---} Although our results are valid for any vortex, in a condensate, only single quantized vortices ($|w| = 1$) are stable~\cite{Fetter09}. When the condensate is spinning faster, higher vortices decay into several single quantized vortices. One can however obtain a stable condensate with a lattice of vortices. The linear character of the long range effect described by \eq{WKBaction} shows that the hydrodynamic Aharonov--Bohm effect cumulates and is only sensitive to the total circulation. More precisely, we consider a lattice of vortices located at $\vec{r_j}$ and of associated  Aharonov--Bohm parameter $\alpha_j$. The phase of the Aharonov--Bohm mode \eqref{WKBmode} reads 
\be
S(r,\theta) = \vec k \cdot \vec r - \sum_j \alpha_j \theta_j, 
\ee
where $\theta_j \in ]0, 2\pi[$ is the angle between $\vec r- \vec{r_j}$ and $\vec k$. At large $r$, the $\theta_j$'s become the same, showing that the total geometric phase is given by the sum $2\pi \sum \al_j$. In Fig.~\ref{front-fig}, we plotted an example for a lattice consisting of $25$ vortices with alined (left column) and alternating (right column) vortices. \\

%
%
\emph{Outlook and summary.---} The observation of solitonic vortex formation, as for example created by the Kibble--Zureck mechanism~\cite{Zurek:1985aa}, play an important role to further advance our understanding of quantum turbulence. The hydrodynamic Aharonov--Bohm effect presented above offers an alternative method to the homodyne detection method~\cite{Chevy01,Donadello14} for quantum vorticity. Within the homodyne detection method the condensate is splitted into two parts. After a period of free expansion the two parts interfere and from the interference pattern one can directly read off the dislocations\,/anti-dislocations of the fringes which correspond to the vortices in the initial condensate~\cite{Donadello14}. Based on the analysis above it is possible to probe the vorticity, i.e.~winding numbers and orientation of vortices, in BECs via the dislocation and discontinuity of phase- \emph{or} density-fluctuations generated by a plane wave propagating through the condensate. The fact that the dislocation is reaching its maximum value --- the number of new wave fronts emerging from a vortex is equal to its winding number --- for $\xi k \gg 1$ (i.e.~high frequency excitations) implies that hydrodynamic Aharonov--Bohm can be utilized as a probe for quantum vorticity. The experimental advantages, when compared to the homodyne technique, are that the hydrodynamic Aharonov--Bohm effect can be observed directly as matter waves in the condensate. 

The detection of the hydrodynamic Aharonov--Bohm effect seems to be within experimental reach. Recent works have shown very promising results, in both the capacity to excite~\cite{Katz:2004aa} and image~\cite{Wilson:2015aa,Sherson:2010aa} plane waves in trapped condensates. When probing quantum vorticity with sub-healing wavelengths, deviations from the predicted pattern are expected, that might reveal details of the core of the quantum vortex. \\

\begin{acknowledgments}
We are particularly thankful for crucial discussions on the context of this paper with Joerg Schmiedmayer, Fedja Orucevic, Vitor Cardoso, Peter Kruger, and Jean Dalibard.
S.W. acknowledges financial support provided under the Royal Society University Research Fellow, the Nottingham Advanced Research Fellow and the Royal Society Project grants. 
This research was supported in part by the Perimeter Institute for
Theoretical Physics. Research at Perimeter Institute is supported by
the Government of Canada through Industry Canada and by the Province
of Ontario through the Ministry of Economic Development $\&$
Innovation.
\end{acknowledgments}

\providecommand{\href}[2]{#2}\begingroup\raggedright\endgroup

\end{document}